The Mattig Expression for a d-dimensional Gauss Bonnet FRWL Cosmology

Keith Andrew[1,2], Eric Steinfelds[1], Nick Zolman[1,3]

[1]Department of Physics and Astronomy
[2]Institute for Astrophysics and Space Science
Western Kentucky University
Bowling Green, KY 42101
[3]Department of Physics, Mathematics, and Astronomy
California Institute of Technology
Pasadena, California, 91125

Abstract

Here we study the form of the Mattig equation applied in a cosmological setting for spacetime metric gravity models described by the Gauss-Bonnet action. We start with expressing the Mattig relation for cosmological magnitudes in terms of standard metric functions and redshift values. Then we present the Gauss-Bonnet field equations and the associated limits for special solutions in an arbitrary number of dimensions. These solutions are used to rewrite the Mattig relation with correction terms from the Gauss-Bonnet contributions for the case where the Gauss-Bonnet scale factor can be directly used to find the distance modulus and for the case where the Gauss-Bonnet field equations can be expressed as a small high z perturbation on the standard Einstein field equations. As a result we can express the perturbative distance modulus, which includes the apparent magnitude, as an additive correction to the standard distance modulus. This results in a small shift in the apparent magnitude of a high z object which gives a contribution depending on the Gauss-Bonnet coupling, spacetime dimension and the cosmological constant.

I.     Introduction

We examine a generalization of the Mattig luminosity distance relation[1] to a Gauss-Bonnet gravitational model in d-dimensions. Improvements in the cosmological distance ladder are ongoing and continue to add to our overall understanding of large scale cosmic structure. Measurements and constraints of cosmological distance scales[2,3] play a role in determining the nature of baryonic acoustic oscillations in the CMB anisotropies[4], from data collected by the Wiggle Z Dark Energy Survey[5], from gamma-ray bursts[6], from the Sunyaev-Zel'dovich effect using Chandra and X-ray sources[7], from gravitational waves from binary neutron star pair coalescences [8], from using quasars as standard clocks[9], looking at the motion of Ly α absorbers[10], from looking for the Sandage-Loeb effect[11], and constraints from the Hubble space telescope[12]. In deriving expressions for a redshift function fundamental cosmological parameters and detailed structure effects enter from HII gas effects[13,14], the low frequency 21 cm line[15] that can involve the cosmological constant, deceleration[16], radiation pressure[17], the intergalactic medium[18], with an influence from TeV Blazars[19], and details of intergalactic cosmic inhomogeneities along a path with many holes[20]. These may be treated as voids[21,22] and the overall distribution of voids can be modeled in terms of a probability distribution[23] with a z dependence[24] and expressed in terms of a partition function or expressed as an integral equation in terms of the Fox H function[25]. There are other more complex cosmological deviations in smoothness[26] which may be treated as perturbations which can be interpreted and modeled as dark energy[27] or quintessence[28] showing the importance of having a strong connection between realistic cosmologies and observations. A number of extensions of General Relativity have also been considered that treat the spacetime manifold M and metric g as the fundamental ingredients for a gravitational interaction: (M, g), where g is constrained by the generalized Einstein field equations and M is characterized by the Riemannian curvature tensor. Many of these theories are inspired by turning to the fundamental ideas of string theory[29] or loop quantum gravity[30] and often effective theories[31] which can be used to account for dark energy observations. In this context geodesic structure[32] provides a tool for investigating cosmological scales, especially null geodesics with their link to causal structure[33].

For interacting theories that involve a scalar field one must select the Einstein or Jordan frame which are conformally related[34], here we use the Jordan frame throughout. Particle interactions in the spacetime manifold are given by the geodesic equations and the geodesic deviation equations for timelike, spacelike and null trajectories which give the generalized forms of the Raychaudhuri[35] and Mattig equations to describe geodesic congruences and distances. Of special interest are the Mattig and Etherington expressions which hold a central role as being amongst the most fundamental expressions that closely link observed magnitudes and redshifts with the basic metric structure of spacetime. The Etherington Reciprocity Law[36], also known as the Cosmological Distance-Duality Relationship[37], is valid for any cosmological model where photons travel on null geodesics in a Riemannian or pseudo-Riemannian manifold. It is a *purely geometric* relation between null geodesics connecting the source to the observer and can be expressed as a fundamental constant as a test of gravitational theories with a unit value given as:

$$\text{Distance} - \text{Duality}: \quad \frac{D_L(z)}{D_A(z)(1+z)^2} = 1$$
$$D_L: \text{luminosity} - \text{distance}$$
$$D_A: \text{angular diameter distance}$$
(1)

for the luminosity and angular diameter distance diameters. In its simplest form it is an expression for null geodesics stating that observers at rest in a static spacetime observe objects of the same size subtending the same solid angle. The Etherington result can also be stated in terms of the emitted temperature of an object, $T_E$, which is related to the received temperature, $T_R$, of the object by: $T_E = (1+z) T_R$, these are basic results for null rays in any cosmology and can be used as a test for any modified dark energy cosmological model. As this theorem is valid for all cosmologies, it has been recently suggested that it can be a powerful tool for testing nonstandard cosmological models, as well as being capable of distinguishing between various models of dark energy or to search for exotic physics[38] by comparing distances measured by x-ray standard candles[39] and standard intrinsic dimensions.

The Mattig relation is not quite as fundamental as the Etherington relation but still serves as a critical connection between cosmological theory and null rays. For a known cosmological solution the Mattig equation can be found by directly integrating the metric coefficients along a null curve connecting the object to the observer. It is valuable to have the distance luminosity function for a given cosmological model expressed in terms of the observables. In general once a cosmological model is assumed then one can define other, model dependent, distances like the commoving distance, proper distance, interval distance, geodesic distance, absolute distance, etc[40]. These are all, however, different forms of line element separations ($ds^2$), whose theoretically-defined expressions are entirely dependent on the spacetime geometry and the particular solution of Einstein's field equations. In the event that spacetime has unusual behavior- not Hausdorff[41], or like a fractal[42,43] or with a holographic structure [44] in nature, greater care is needed in defining a consistent distance measure for light rays from which cosmological distances can be found. The particular expressions change according to the cosmological model and, hence, they are not observationally defined distances, although they do play an important theoretical role in cosmology.[45] One of the well-studied extensions of the Einstein field equations are the Gauss-Bonnet modifications and modified Gauss-Bonnet theories[46]. These string inspired corrections[47] provide insights into methods for treating dark energy[48], extensions to brane cosmology[49], extensions to d-dimensions[50], with inflation[51] and compactification[52], leading to new insights and tests of large scale structure[53]. Here we use the Gauss-Bonnet field equations to get a model dependent correction to the Mattig cosmological distance relation. Such models have been the subject of astrophysical tests and continue to be used to gain insight into modifications to General Relativity[54].

In the next section we review the Mattig relation in a standard FLRW cosmology using the direct integration of the known metric terms, in section 3 we then set up the d-dimensional Gauss-Bonnet solutions and notation, then in section 4 we integrate these solutions and find expressions for the Mattig relation by direct integration of the Gauss Bonnet expansion factor and then as a perturbative addition to the Einstein field equations.

## II.     Mattig Relations in FLRW cosmology

Consider the standard FLRW metric ansatz using the conformal time transformation with scale factor a(t):

$$ds^2 = dt^2 - a^2(t)\left[\frac{dr^2}{1-kr^2} + r^2\left(d\theta^2 + \sin^2 d\phi^2\right)\right] = a^2(t)\left[d\eta^2 - \frac{dr^2}{1-kr^2} + r^2 d\Omega^2\right]. \quad (2)$$

$$where \quad dt = a(t)d\eta$$

The resulting equations with a cosmological constant are given by

$$R_{\mu\nu} - \frac{1}{2} g_{\mu\nu} R + g_{\mu\nu} \Lambda = \kappa T_{\mu\nu} \quad T^{\mu\nu} = [\rho, p, p, p]$$

$$\left(\frac{\dot{a}}{a}\right)^2 + \frac{k}{a^2} - \frac{\Lambda}{3} = \frac{\rho}{3} \quad (3)$$

$$2\frac{\ddot{a}}{a} + \left(\frac{\dot{a}}{a}\right)^2 + \frac{k}{a^2} - \Lambda = -p$$

These equations can be expressed in terms of the Hubble parameter, H, deceleration parameter, q, and dimensionless ratios- the density parameters, resulting from matter, radiation, curvature, k, and the cosmological constant, $\Lambda$, compared to the critical density value, $\rho_c$, for an equation of state (EoS) relating pressure to density and using the conservation equations to relate the density to scale parameter, using the following definitions:

$$H = \frac{\dot{a}}{a} \quad q = -\frac{\ddot{a}a}{\dot{a}^2} = -\frac{\ddot{a}}{aH^2} \quad \rho_c = \frac{3H^2}{8\pi}$$

$$EoS: p_i = w_i \rho_i$$

$$n_i = 3(1 + w_i) \quad (4)$$

$$Conservation: \nabla_\mu T^{\mu\nu} = 0 \rightarrow \rho_i \propto a^{-n_i}$$

$$\Omega_M = \frac{\rho}{\rho_c} \quad \Omega_R = \frac{\rho_R}{\rho_c} \quad \Omega_\Lambda = \frac{\rho_\Lambda}{\rho_c}$$

From the FRW Eqs.(2) the Hubble parameter can be expressed in terms of the dimensionless ratios for matter, curvature and cosmological constant, measured from the current epoch $t=t_0$ using the equations of state $p_i = w_i \rho_i$ for each contributing factor to the density: matter (i=1), radiation (i=2) and cosmological constant (i=3) that relates pressure to density with redshift factor by (z+1). Then these expressions can be expressed directly in terms of the resulting densities, $\rho_i$, or in terms of the dimensionless ratios, $\Omega_i$, giving the simplified expressions:

$$\Omega_i = \frac{\rho_i}{\rho_c} \quad \Omega_1 = \Omega_M = \frac{\rho}{\rho_c} \quad \Omega_2 = \Omega_R = \frac{\rho_R}{\rho_c} \quad \Omega_3 = \Omega_\Lambda = \frac{\Lambda}{\rho_c}$$

$$\rho_1 = \rho_m \quad \rho_2 = 0 \quad \rho_2 = \rho_k = \frac{k}{2} \quad \rho_3 = \rho_\Lambda = \frac{\Lambda}{3} \quad (5)$$

$$\frac{a_0}{a(t)} = 1 + z \quad Constraint: \sum_j \Omega_j - 1 = \Omega_k = \frac{-k}{\rho_c}$$

where the redshift is related to the ratio of the scale factors for any epoch, the radiation density is zero for times well beyond the radiation epoch, the curvature constant contributes to the overall density in a fashion similar to matter and radiation, and the current observational constraint[55] on the total dimensionless mass energy demands the sum of all terms is unity and observationally k~0.

Combining the FRWL equations with the expansion and scale factor equations with the metric expression for a null ray the resulting null distance is

$$ds^2 = 0 = dt^2 - \frac{a^2 dr^2}{1-kr^2} \quad da = \frac{a_0 dz}{(1+z)^2}$$

$$\left(\frac{\dot{a}}{a}\right)^2 = \Omega_M H_o^2 (1+z)^3 + \Omega_k H_o^2 (1+z)^2 + \Omega_\Lambda H_o^2 = H^2$$

$$H(z) = H_0 \sqrt{\Omega_{M1}(1+z)^3 + \Omega_{k0}(1+z)^2 + \Omega_{\Lambda 0}} = \frac{\dot{a}}{a}$$

$$\int_0^r \frac{dr'}{\sqrt{1-kr'^2}} = \int \frac{dt}{a(t)} = \int \frac{da}{\dot{a}a} = \int \frac{-dz}{\dot{a}(1+z)} = \arcsin r = \chi$$

(6)

Using the FRW equation to solve for $\dot{a}$ in terms of H(z) gives the standard form

$$\arcsin r = \frac{1}{H_0} \int \frac{dz}{\sqrt{\Omega_{M0}(1+z)^3 + \Omega_{k0}(1+z)^2 + \Omega_{\Lambda 0}}} \qquad (7)$$

Eq.(7) can be solved analytically for the special case originally worked out by Mattig corresponding to a matter dominated universe, and for $q_0 > 0$ then

$$r = \left(\frac{1}{a_o H_o}\right)\left[\frac{q_o z + (q_o - 1)(-1 + \sqrt{1+2q_o z})}{q_o^2(1+z)}\right] \qquad (8)$$

As shown by Terrell[56] the luminosity distance and absolute magnitudes can then be written for each case of the deceleration parameter and is often expressed as

$$D_L = a_o r_{Mattig}(1+z) = \frac{c}{H_o}\left[z + \frac{z^2(1-q_o)}{1+q_o z + \sqrt{1+2q_o z}}\right] \quad q_0 \geq 0$$

(9)

In terms of the distance modulus defined as the difference between the absolute and apparent magnitudes for luminosity distance $D_L$ for a value of r given in Eq.(8) we have

$$M_{absolute} = m_{apparent} - 5\log(D_L)_{Mpc}$$
$$\mu = m - M = 5(\log_{10}[a_o r(1+z)] - 1) \qquad (10)$$

Knowing the redshift z, the current value of the scale parameter $a_o$ and the Mattig result from Eq.(10) we can calculate the distance modulus or with the apparent magnitude the absolute magnitude of the object. This approach can also be used to explore the magnitude shift that arises from extensions of the standard version of the Einstein equations. This is especially useful for small changes that arise in modifications like the Gauss-Bonnet extension which can be treated in a perturbative fashion. This can be approached by solving the modified geodesic equations first or by following this procedure starting with the Gauss-Bonnet modified Einstein equations.

### III.  Gauss-Bonnet Solutions

One extension of the Einstein field equations that is motivated by the low energy limit of string theory and gives a second order system of differential equations is the Gauss Bonnet extension. The Gauss Bonnet extension includes quadratic curvature terms and can be extended to an arbitrary number of dimensions in a natural way[57]. Current experiments do provide constraints on the Gauss Bonnet couplings[58][59] and the maximum current size (of extra dimensions) to be ~ 100 μm[60][61][62]. An idea that has been extensively studied is that these extra dimensions were once large but underwent dynamical compactification as the usual three spatial dimensions grew. This has been explored by Paul and Murkherjee[63] and Mohammedi[64] in an analytical way that can be used as the basis for extending their solutions[65]. It is also widely thought that Einstein gravity is only a low energy effective field theory which requires modification at higher energy. A detailed and insightful review of this connection was given by Boulware and Deser[66]. One possible modification of the Einstein-Hilbert action comes from adding additional Lovelock terms[67]. This modification is attractive because it yields second-order, divergence free field equations as one would demand from a generally covariant theory of gravitation[68]. One may formulate Lovelock gravity theory as an expansion in powers of the curvature to obtain a zeroth-order constant term or cosmological constant, a first-order term that gives the usual scalar curvature which yields Einstein gravity, and a second-order term that is known as the Gauss-Bonnet term plus higher order terms. Here we incorporate both of these ideas: compactification of the higher dimensions and the addition of a Gauss-Bonnet term to the action.

We consider a dynamical compactification of a D-dimensional manifold to a maximally symmetric manifold of dimension d and an expanding FRW spacetime of dimension 4 where we have modified the Einstein-Hilbert action by including a Gauss-Bonnet term. This Gauss-Bonnet term can be interpreted as being a first order correction from string theory or simply a modification of Einstein gravity

$$S = \int d^D x \sqrt{-g} \left( R - 2\lambda - \varepsilon\, G_{GB} \right)$$
$$G_{GB} = R_{ABCD} R^{ABCD} - 4 R_{AB} R^{AB} + R^2 \quad .$$

(11)

We use the metric ansatz in D dimensions

$$ds^2 = -dt^2 + a^2(t)\left[\frac{dr^2}{1-Kr^2} + r^2(d\theta^2 + \sin^2 d\phi^2)\right] + b^2(t)\gamma_{mn}(y)dy^m dy^n \qquad (12)$$

$$R_{abcd} = k(\gamma_{ac}\gamma_{bd} - \gamma_{ad}\gamma_{bc}) \qquad (13)$$

Where the extra dimensions are related to the scale factor by

$$b(t) \sim \frac{1}{a^n(t)} \quad n > 0 \qquad (14)$$

This gives the following field equations

$$\begin{aligned}
\frac{\rho}{2\kappa} &= \eta_1 \frac{\dot{a}^2}{a^2} + \varepsilon\xi_1 \frac{\dot{a}^4}{a^4} \\
\frac{p}{2\kappa} &= \left[\eta_2 \frac{\ddot{a}}{a} + (\eta_1 - \eta_2^2)\frac{\dot{a}^2}{a^2}\right] + \varepsilon\left[\frac{1}{3}(\xi_1 - dn\xi_2)\frac{\dot{a}^4}{a^4} + \xi_2 \frac{\ddot{a}\dot{a}^2}{a^3}\right] \\
\frac{p_d}{2\kappa} &= \frac{1}{dn}\left[(2\eta_1 + 3\eta_2)\frac{\ddot{a}}{a} + [2(2\eta_1 + 3\eta_2) - dn(\eta_1 + 3\eta_2)]\frac{\dot{a}^2}{a^2}\right] \\
&\quad - \varepsilon\left[(\xi_1 + \xi_2)\frac{\dot{a}^4}{a^4} - \frac{1}{dn}(4\xi_1 + 3\xi_2)\frac{\ddot{a}\dot{a}^2}{a^2}\right]
\end{aligned} \qquad (15)$$

where we define

$$\xi_1 = -dn\left[(d-1)n\left[\frac{1}{2}(d-2)n[(d-3)n-12]+18\right]-12\right] \qquad (16)$$
$$\xi_2 = -2dn[(d-1)n[(d-2)n-6]+6]$$

$$\eta_1 = \frac{1}{2}[6 + dn(dn - n - 6)] \qquad (17)$$
$$\eta_2 = (dn - 2)$$

where the extra dimensions are defined to be maximally symmetric such that the Riemann tensor for γ has the form $R_{abcd} = k(\gamma_{ac}\gamma_{bd} - \gamma_{ad}\gamma_{bc})$. In agreement with current observations we will consider the usual 3 spatial dimensions to be flat (K = 0) and also demand that the extra dimensions be flat (k = 0). A perturbative solution for a(t) is then given by

$$w \neq -1$$

$$a(t) \approx \beta t^{2/3(1+w)}\left[1+\varepsilon\frac{\xi_1}{2\eta_1}\left[\frac{2}{3(1+w)}\right]^3\frac{1}{t^2}\right]$$

$$\beta = \left[\left[\frac{3}{2}(1+w)\right]^2\frac{\rho_o}{2\kappa\eta_1}\right]^{1/3(1+w)}$$
(18)

This solution is in the form given as a standard FRWL term plus a perturbative Gauss Bonnet correction factor. Eq.(18) can now be combined with the Mattig relation from Eq.(6) and Eq.(8) to give the correction to the magnitude and distance modulus for a d-dimensional Gauss-Bonnet gravity model cosmology.

### IV. Generalized Gauss Bonnet Mattig Equation

Here we calculate the Mattig relation by the direct substitution of Eq.(18) into Eq.(6) and by treating the Gauss Bonnet term as a small perturbation on the FRWL equations. Starting with the time dependent scale factor from Eq.(18) we have (for $w \neq -1$)

$$\chi = \arcsin r = \int_t^{t_o}\frac{dt'}{a(t')} = \frac{1}{\beta}\int_t^{t_o}\frac{t'^{-2/3(1+w)}dt'}{\left[1+\varepsilon\frac{\xi_1}{2\eta_1}\left(\frac{2}{3(1+w)}\right)^3\frac{1}{t'^2}\right]} = \frac{1}{\beta}\int_t^{t_o}\frac{t'^{-2/3(1+w)+2}dt'}{\left[t'^2+\varepsilon\frac{\xi_1}{2\eta_1}\left(\frac{2}{3(1+w)}\right)^3\right]}$$
(19)

Which can be simplified to

$$\chi = \frac{1}{\beta}\int_t^{t_o}\frac{(t')^{2-c_1}}{(t')^2+\varepsilon c_2}dt'$$

$$\chi = \frac{t^{1-c_1}}{\beta(c_1-1)}\left[{}_2F_1\left(1,\frac{1-c_1}{2};\frac{3-c_1}{2};-\frac{t^2}{\varepsilon c_2}\right)-1\right].$$
(20)

$$c_1 = \frac{2}{3(1+w)} \qquad c_2 = \frac{\xi_1}{2\eta_1}\left(\frac{2}{3(1+w)}\right)^3$$

Where the hypergeometric function is defined in terms of Pochhammer symbols as

$${}_2F_1(a;b;c;z) = \frac{[a]_m[b]_m}{[c]_m m!}z^m$$

$$[a]_m = a(a+1)(a+2)....(a+m-1) = \frac{\Gamma(a+m)}{\Gamma(a)}$$
(21)

Expanding about the origin for small corrections gives

$$r \approx \frac{1}{\beta}\left(\frac{t^{1-c_1}}{1-c_1} + \varepsilon \frac{c_2 t^{-1-c_1}}{(1+c_1)}\right). \tag{22}$$

The resulting luminosity distance and distance modulus expressions are

$$\begin{aligned}
\mu &= \mu_{Mattig} + \varepsilon \mu_{GB} \\
\mu_{Mattig} &= 5\log\left(\frac{t^{1-c_1}}{\beta(1-c_1)}\right) - 5 \\
\mu_{GB} &= \frac{5c_2(1-c_1)t^{-2}}{(1+c_1)}
\end{aligned} \tag{23}$$

The resulting distance modulus includes a small shift due to the Gauss-Bonnet terms giving a small change in the apparent magnitudes. This expression is not in the standard Mattig form due to the explicit independent variable which is defined through Eq.(6) and the relationship between the scale factor a(t) and z.

This dependence can be removed by alternatively starting with the field equations from Eqs.(15) and the definitions from Eqs.(4) to rewrite the field equations as a reduced quartic equation

$$\varepsilon \xi_1 \frac{\dot{a}^4}{a^4} + \eta_1 \frac{\dot{a}^2}{a^2} - H_o^2\left(\Omega_{om}(1+z)^3 + \Omega_{ok}(1+z)^2 + \Omega_{o\Lambda}\right) = 0 \tag{24}$$

Solving for the rate of change of scale factor and taking the root indicating a small perturbative change for the Gauss-Bonnet term gives:

$$\dot{a} = H_o\left(\Omega_{om}(1+z)^3 + \Omega_{ok}(1+z)^2 + \Omega_{o\Lambda}\right)^{1/2}\left[1 - \frac{\varepsilon \xi_1}{2\eta_1 a^4} H_o^2\left(\Omega_{om}(1+z)^3 + \Omega_{ok}(1+z)^2 + \Omega_{o\Lambda}\right)\right] \tag{25}$$

$$\varepsilon \xi_1 \frac{\dot{a}^4}{a^4} + \eta_1 \frac{\dot{a}^2}{a^2} - H_o^2\left(\Omega_{om}(1+z)^3 + \Omega_{ok}(1+z)^2 + \Omega_{o\Lambda}\right) = 0 \tag{26}$$

Distributing the leading factor and changing variables to the redshift while integrating both sides of Eq.(26) gives

$$\arcsin r = \int \frac{dz\left[1 + \frac{\varepsilon \xi_1}{2\eta_1 a_o^4} H_o^2\left(\Omega_{om}(1+z)^7 + \Omega_{ok}(1+z)^6 + \Omega_{o\Lambda}(1+z)^4\right)\right]}{H_o(1+z)\left(\Omega_{om}(1+z)^3 + \Omega_{ok}(1+z)^2 + \Omega_{o\Lambda}\right)^{1/2}} \quad (27)$$

Which can be separated into a classical Mattig term and a perturbative Gauss-Bonnet correction term

$$\arcsin r = \int \frac{dz}{H_o(1+z)\left(\Omega_{om}(1+z)^3 + \Omega_{ok}(1+z)^2 + \Omega_{o\Lambda}\right)^{1/2}} + \varepsilon \int \frac{dz\left[\frac{\xi_1}{2\eta_1 a_o^4} H_o^2\left(\Omega_{om}(1+z)^7 + \Omega_{ok}(1+z)^6 + \Omega_{o\Lambda}(1+z)^4\right)\right]}{H_o(1+z)\left(\Omega_{om}(1+z)^3 + \Omega_{ok}(1+z)^2 + \Omega_{o\Lambda}\right)^{1/2}} \quad (28)$$

$$\arcsin r = I_{Mattig} + \varepsilon\, I_{GB} = I_{Mattig} + \varepsilon(I_{om} + I_{ok} + I_{o\Lambda})$$

The first term is the standard Mattig term and the second term is the Gauss-Bonnet contribution. For small z values and the known values for the baryonic, curvature and dark energy contributions where the curvature term is observed to be consistent with a zero value, the baryonic term is small compared to the dark energy term giving three integrals, in terms of the independent redshift variable, that can be approximated as

$$I_{om} = \int \frac{dz\left[\Omega_{om}(1+z)^7\right]}{(1+z)\left(\Omega_{om}(1+z)^3 + \Omega_{ok}(1+z)^2 + \Omega_{o\Lambda}\right)^{1/2}} \rightarrow \frac{\Omega_{om}(1+z)^7}{7\Omega_{o\Lambda}^{1/2}} \quad (29)$$

$$I_{ok} = \int \frac{dz\left[\Omega_{ok}(1+z)^6\right]}{(1+z)\left(\Omega_{om}(1+z)^3 + \Omega_{ok}(1+z)^2 + \Omega_{o\Lambda}\right)^{1/2}} \rightarrow 0 \quad (30)$$

$$I_{o\Lambda} = \int \frac{dz\left[\Omega_{o\Lambda}(1+z)^4\right]}{(1+z)\left(\Omega_{om}(1+z)^3 + \Omega_{ok}(1+z)^2 + \Omega_{o\Lambda}\right)^{1/2}} \rightarrow \frac{\Omega_{o\Lambda}^{1/2}}{4}(1+z)^4 \quad (31)$$

Using the magnitude luminosity relation the Gauss-Bonnet correction can be expressed as:

$$r(z) = r_{Mattig} + \varepsilon\, r_{GB} = r_{Mattig} + \varepsilon \frac{\xi_1(1+z)^4 H_o \Omega_{o\Lambda}^{1/2}}{8\eta_1 a_o^4}\left[1 + \frac{4\Omega_{om}}{7\Omega_{o\Lambda}}(1+z)^3\right] \quad (32)$$

The luminosity distance and the distance modulus have the form

$$D_L = R_o(1+z)[r_{Mattig} + \varepsilon\, r_{GB}]$$

$$\mu = 5\log_{10}(R_o(1+z)[r_{Mattig} + \varepsilon\, r_{GB}]) - 5 = 5\log_{10}(R_o(1+z) r_{Mattig}) + 5\log_{10}\left(1+\varepsilon\frac{r_{GB}}{r_{Mattig}}\right) - 5$$

$$\mu = \mu_{Mattig} + 5\varepsilon\frac{r_{GB}}{r_{Mattig}} = \mu_{Mattig} + \varepsilon\mu_{GB}$$

$$\mu_{GB} = \frac{5\xi_1(1+z)^4 H_o \Omega_{o\Lambda}^{1/2}}{8\eta_1 r_{Mattig} a_o^4}\left(1 + \frac{4\Omega_{om}}{7\Omega_{o\Lambda}}(1+z)^3\right)$$

(33)

where the small GB correction simplifies the logarithmic term to the first term approximation from the series expansion of the log function. The advantage to this expression is that the Gauss-Bonnet term is a small additive correction to the standard Mattig distance modulus which can be expressed as a small correction in magnitude. Depending on the details of the GB model- the strength of the coupling, the values of n and d, the distance modulus will shift, which is especially significant for high z values.

## V.    Conclusions

We have studied the Gauss-Bonnet contribution to the Mattig relation in a FRWL setting with general values of n and d. We find there is a correction to the Mattig relation which causes a shift in the luminosity distance and the distance modulus. The change in the distance modulus yields a change in the apparent magnitude of the object, although the change can be positive or negative depending upon the coupling we find the decrease in magnitude to be interesting. Here such a decrease results from both the Gauss-Bonnet terms and the cosmological constant contribution. In addition such a decrease continues to grow with increasing values of z causing a small dimming in magnitude for large distances.